\documentclass{article} 

\usepackage{cite}
\usepackage{hyperref}
\usepackage[pdftex]{graphicx}
\DeclareGraphicsExtensions{.pdf,.jpeg,.png}
\usepackage[caption=false,font=footnotesize]{subfig}
\usepackage{url}
\usepackage{listings}
\begin{document}

\title{A Security Analysis of \\ CheriBSD and Morello Linux}

\author{Dariy Guzairov, Alex Potanin, Stephen Kell\footnote{King's College London}, Alwen Tiu\\Australian National University}

\maketitle

\begin{abstract}
Memory corruption attacks have been prevalent in software for a long time. Some mitigation strategies against these attacks do exist~\cite{memory_corruption_mitigations_2018}, but they are not as far-reaching or as efficient as the CHERI architecture~\cite{cheri_c18n_2023}. CHERI uses capabilities to restrict pointers to certain regions of memory and with certain access restrictions. These capabilities are also used to implement ``compartmentalisation'': dividing a binary into smaller components with limited privilege, while adhering to the principle of least privilege. However, while this architecture successfully mitigates memory corruption attacks, the compartmentalisation mechanisms in place are less effective in containing malicious code to a separate compartment. This paper details four ways to bypass compartmentalisation, with a focus on Linux and BSD operating systems ported to this architecture. We find that although compartmentalisation is implemented in these two operating systems, simple bugs and attacks can still allow malicious code to bypass it. We conclude with mitigation measures to prevent these attacks, a proof-of-concept demonstrating their use, and recommendations for further securing Linux and BSD against unknown attacks.
\end{abstract}

\section{Introduction}

Hardware-enforced security mitigations are becoming more popular in the security research community~\cite{yu_capstone_2023,cheri_c18n_2023}. CHERI project in particular is widely promoted by the UK government as the ``new standards in secure system design''~\cite{CHERI_UKGOV_2025}. 
A core feature of CHERI architecture is a notion of {\em capabilities} attached to a pointer, representing the ranges and the permissions associated with the pointer. This can be seen as a hardware implementation of ``fat pointers''~\cite{nagarakatte2009softbound}. 
Each memory address is 128 bits long, with the first 64 bits used for the usual value and the last 64 bits for the ``capabilities and permissions'' that can be efficiently checked in hardware at runtime. Each capability is additionally augmented with an out-of-band bit denoting its validity. Recently, this architecture was implemented in the ARM Morello prototype boards. This is the implementation on which we conducted our experiments on. The short argument for memory safety under this enhanced architecture is that by compiling against the CHERI architecture, libraries and using the established C and C++ development toolchain, one can gain memory safety guarantees comparable to Rust or other more complex systems with relatively small programmer productivity or performance penalties~\cite{cheriDescriptionOverview}.


\textbf{Compartmentalisation} is one of the core advanced concepts of the CHERI architecture, with the aim of separating code and data of \textit{trusted} compartments from \textit{untrusted} compartments. This is mainly achieved through spatial and temporal separation of memory regions, enforced through the use of CHERI capabilities, which prevent libraries from accessing each other's code and data. 
With the UK government encouraging developers to utilise and develop on the CHERI architecture~\cite{CHERI_UKGOV_2025}, we aim to show that, while this architecture provides a rich set of hardware supported primitives to prevent memory corruption bugs, there are still issues in  adapting these primitives to provide memory safety in real-world software systems, most notably in providing secure compartmentalisation of software systems and libraries.  We discuss two platforms, Morello Linux and CheriBSD, and show how malicious code can bypass compartmentalization, leading to a final proof-of-concept in which a malicious library retrieves a private key from the main binary.

We use the same threat model as Gutstein~\cite{CheriSecAnalysis}, that of an \textit{attacker that has arbitrary code execution within an unprivileged library of
some process and aims to gain code execution or access to secrets in a more privileged
library}. Such an attack scenario is plausible in the case where a malicious library is used within an application, e.g., through a supply chain compromise~\cite{ensia_supplyChain_2021,Ladisa2023Taxonomy}. We consider only attacks that compromise memory safety at the user space, so the malicous code is not assumed to run at a system privilege. We also consider as out of scope attacks that exploit direct system calls (such as calling \lstinline{execve()} to run arbitrary binary) and attacks that exploit (pseudo-)file systems (such as \lstinline{/proc/self/mem} proxy for a process virtual memory in linux),  
but the malicious code can call functions in common libraries such as \lstinline{libc}. With this threat model, we aim to show how a malicious library is able to break compartmentalisation and access secrets in a more privileged compartment or library. We show that the malicious actor has escaped the compartment by reading or writing to a variable in the main binary, such as a private key or a flag. 

CHERI provides a set of APIs to interface with the hardware instructions to manipulate capabilities. We assume these hardware primitives are secure, so the only to obtain capabilities is to manipulate existing ones. Our exploration thus focuses on examining ways in which capabilities can be (unintentionally) leaked in the presence of compartmentalisation. We show that there are broadly two ways in which capabilities can be leaked (in the Morello Linux and CheriBSD implementations on ARM Morello boards): through exploiting an implementation flaw in dynamic linkers and through a {\em memory scavenging} technique. More specifically, our contributions are as follows.
\begin{itemize}
\item {\em Dynamic linker exploit.} We show that in both Morello Linux and CheriBSD, certain library calls that interface with the dynamic linker (such as \lstinline{dlopen()} -- that is supported in both platforms) allows the malicious code to obtain capabilities to all loaded libraries and the main binary. 
\item {\em Memory scavenging.} We show that capabilities that are inadvertently left on the stack or heap objects, e.g., because a function does not zero a particular buffer prior to exiting, or due to the heap allocator not implementing memory zeroing when freeing unused objects, can potentially be accessed by the malicious code through systematic scavenging. Depending on the application logic, these left-over capabilities can be leveraged to obtain further capabilities, through a recursive scan of the stack or the heap. 

\end{itemize}
While both attack surfaces are anticipated in the literature~\cite{Cheri_Security_Analysis_2022,Morello-Cerise-paper}, as far as we know, this is the first demonstration that such exploits are feasible and practical.  

In summary, our contribution lies in providing a technique to find every reachable capability from an initial capability on both CheriBSD and Morello Linux~\ref{sec:RecursiveScanning}. 
We additionally discuss any mitigations that exist, and proposed mitigations against these attacks.

The proof of concept code for the exploits detailed in this paper can be found in an anonymized  repository.\footnote{\url{https://gitfront.io/r/anonuser2025/JUZMJC7m7zwa/cheri-security/}}


\section{Background}
\label{sec:Background}

\subsection{CHERI Architecture}

While the CHERI architecture is similar to many other standard architectures, capabilities drastically change what can be done. 

We begin with the CHERI architecture but do not discuss the hardware or architecture itself. As stated before, this is a new computer architecture proposed and developed by ARM in collaboration with the University of Cambridge~\cite{cheri_ARM_2025}. The main feature of this architecture is the introduction of 128-bit registers, the concept of capabilities, and compartmentalisation within a binary program. These capabilities are used to enforce the \textit{principle of least privilege}, aiming to prevent the impact of memory corruption vulnerabilities, such as heap and buffer overflows~\cite{CHERI_allocator_2023,hardware_cap_compartmentalization_2023}. The way CHERI implements this principle is through \textit{fine-grained memory protection and scalable software compartmentalisation}~\cite{CHERI_arch_reference_2017}. This is done using 128-bit words, split into a 64-bit pointer, a base, a length, an offset, a validity bit, and a permission mask \cite{CHERI_arch_reference_2017}. The base, length, and offset describe the region of memory the capability can interact with, while the permission mask describes what the capability can do, such as reading or writing. The 64-bit pointer remains the same, allowing reading or writing to the memory that it points to. The validity bit is not stored inline with the capability, as it dictates whether the capability is valid, preventing the forging of capabilities. An example of different capabilities (presented in a ``pretty printed'' format) is shown in Figure~\ref {fig:capability_example}.

\begin{figure}
    \centering
    \begin{lstlisting}[basicstyle=\small\ttfamily,language={[x86masm]Assembler}]
0xfffff5e18314 
[rwRW,0xfffff5e1830c-0xfffff5e18331]

0xaaaad4fb0000 
[rxRE,0xaaaad4fb0000-0xaaaad4fe2000]

0x40195cc0 
[rwRW,0x40195cc0-0x40195d20] (sealed)

0x40152d61 
[rxR,0x40130000-0x40190700] (sentry)

0x40c19000 
[rwRW,0x40c19000-0x40c19010] (invalid)
    \end{lstlisting}
    \caption{Several example Capabilities.}
    \label{fig:capability_example}
\end{figure}

As shown, the 64-bit pointer is the same as it is in C: a memory address. After this pointer, we see the permission bits. In this case, the capabilities are either read and write (\lstinline{rwRW}), or read and execute (\lstinline{rxRE}). The lowercase permissions signify reading and writing standard data, while the uppercase permissions signify reading and writing capabilities. If you attempt to read a capability without these uppercase permission bits, the capabilities you will read will be invalidated. After the permissions is the start address, followed by the end address. The start and end addresses are calculated from the offset, base, and length that a capability holds, and denote what region of memory the capability can reach. If a capability tries to access memory outside of this range, a segmentation fault occurs. Notably, the length of a capability allows for a wide range of use cases. It can be fine-grained, with a precise byte length, or it can be larger, spanning multiple pages. In the brackets, the sealed and sentry designators indicate if a capability is sealed. A sealed capability is immutable and cannot be dereferenced~\cite{cheri_architecture_2023}. This is useful for protecting global data or data passed to a malicious library, or data that you otherwise do not want to change. A sealed entry, or sentry designator, is an executable capability that cannot be modified, but can still be called to run. In essence, it makes sure that the only entry point to that region of code is through this address. Finally, the invalid designator shows us that a capability is invalid. This occurs when the tag bit is set to 0. This happens either when a capability is set to be invalid through code, when a capability is outside of the set range, or if data is read that isn't a capability. Notably, the tag bit is stored separately from the 128-bit capability (using the ECC bits), which prevents malicious code from crafting a capability that points to any address of their choosing. 

As sealing is a powerful primitive that makes capabilities immutable, we describe some of the intricacies that come with sealing. To seal capabilities, a program can use a key to seal a capability, and this capability can only be unsealed with that same key. An attempt to access or to dereference a sealed capability would cause a segmentation fault. One interesting aspect of Morello Linux is the concept of a \textit{global sealer} (which is used in an example program in the official Morello SDK) which serves as the key to seal the private data. If we are able to get the global sealer or any other key used to seal a capability, we can unseal it and access it as usual. 

Additionally, capabilities~\cite{RobustComposition,melicher_et_al:LIPIcs.ECOOP.2017.20,10.1007/978-3-030-02450-5_14,Noble2018,GARIANO2022102878} provide software compartmentalisation through these memory protections, restricting where in memory the code can execute, where the program has read and write access, and other such restrictions. This is in addition to existing restrictions enforced by the memory management unit and page table, such as read-only or executable pages \cite{cheriDescriptionOverview}. 

\subsection{Compartmentalisation}

Compartmentalisation is one key way to confine malicious or untrusted code from trusted code. By isolating key components of software into separate compartments, we can isolate high-risk components, ideally preventing attacks that target high-risk code from being able to access privileged and secure code~\cite{CHERI_arch_reference_2017}. As such, the isolation of untrusted code away from trusted code is a key part of compartmentalisation~\cite{Dynamic_library_compartmentalization_2023}. This technique has been attempted many times in the past; some focused on compartmentalising through software~\cite{Compile_Time_compart_2023}, while others focused on compartmentalising through hardware~\cite{Software_Compartmentalization_2023}. 
CHERI~\cite{cheriDescriptionOverview} provides multiple design patterns for compartmentalisation, with some examples being: \textit{Sandboxing, Assured pipelines, Horizontal compartmentalisation, library compartmentalisation, and Temporal compartmentalisation}. Each of these patterns aims to prevent malicious code from escaping, either by granting the compartment a limited set of rights or by encapsulating components of a larger system within its own compartment. For our attacks, we will focus on bypassing compartmentalisation within a single process, as that is the key security feature that CHERI proposes over traditional sandboxing via Inter-Process Communication (IPC). 

As IPC relies on passing messages between different processes, sandboxing using different processes incurs a high overhead as you need to create a new process for each sandbox, and share data in a page granularity. If you only need to give the sandboxed code a small amount of data, sending a full page is expensive and inefficient \cite{cheriDescriptionOverview}. In contrast, capabilities allow CHERI to compartmentalise within a single process, confining a portion of the code to only access data and code inside a subset of the full memory space. Now, to pass data from one compartment to another, all you need to do is give it a capability that points to the data you want to share. This capability will be confined to only permit access to said data, and in theory, the compartments will remain isolated. As every compartment is in the same address space, there is very little overhead when passing data from one compartment to another. However, this presents the issue of transitive closure, as a leak of a capability presents a way for malicious code to break out of its compartment. This is the issue that we will explore, and how simple information leaks can break compartmentalisation.

\subsection{The Dynamic Linker}

When compiling programs, one could link a program statically or dynamically. Dynamic linking allows a program to avoid linking every library together with the binary. Instead, the dynamic linker is able to load the required libraries at runtime. The main benefit of this is a smaller-sized binary, as the required libraries are loaded automatically by the dynamic linker before program execution has begun, instead of being inlined with the binary. 

To do this, a \textit{global offset table} (GOT) is established in the program's memory, and pointers to library functions are placed there by the dynamic linker. When the binary attempts to call a function inside a library, it jumps to the GOT, which has a placeholder address that points to the dynamic linker. This is shown at the top of figure \ref{fig:linkingExample}. When this placeholder address is called, the dynamic linker finds and substitutes the real address of the function back into the GOT, replacing the placeholder address, and calls the requested function. This is the key mechanism that allows libraries to be loaded into an executable at runtime, rather than at compile-time \cite{dynamic_linker_2024}. In order to do this, the dynamic linker needs to establish and maintain metadata about each of the loaded libraries, such as their base address, name, size, symbol table, and a number of other metadata that is used to map it into the binary, and fill out the GOT to point to the correct functions. As such, the dynamic linker has a lot of permissions and access to the binary, especially on the CHERI architecture. 

\begin{figure}
    \centering
    \begin{lstlisting}[basicstyle=\small\ttfamily,language={[x86masm]Assembler}]
GOT before dynamically loading:
<puts@got.plt>:   0x0000555555555036      
<printf@got.plt>: 0x0000555555555046
<dlopen@got.plt>: 0x0000555555555056

GOT after dynamically loading:
<puts@got.plt>:   0x00007ffff7e2b5a0
<printf@got.plt>: 0x00007ffff7e04900
<dlopen@got.plt>: 0x00007ffff7e3a2a0
    \end{lstlisting}
    \caption{Comparison of a GOT table.}
    \label{fig:linkingExample}
\end{figure}

As dynamic linking is often the default when compiling software, the dynamic linker is ubiquitous and makes for an interesting target for us, especially in the context of compartmentalisation. 

\subsection{Software exploitation}

As our attacks are relatively novel, they do not rely on much of the existing exploitation background. 

The most notable existing attacks and methodologies are \textit{egghunters} and \textit{sandbox escapes}. 

Sandboxes are a concept in software security that aims to isolate untrusted code. This isolation limits what the untrusted code can do, for example, reading or writing to a small subset of files or having access to a subset of functions that the trusted code can access \cite{WebIsolation_2009}. Oftentimes, these sandboxes are child processes and have restrictions placed on what they can do, such as only rendering the page and running JavaScript.

While stack and heap exploitation give malicious actors a means of achieving code execution, sandbox escapes are a means of bypassing a sandbox \cite{PHRACK_sandbox_2021}. Oftentimes, these escapes also rely on bugs such as overflows or logic bugs, which allow the malicious code to escape the sandbox. In our case, compartmentalisation plays a similar role, confining our malicious code to a small section of memory. Due to the CHERI architecture, we cannot use buffer overflows or other memory corruption attacks to escape a sandbox, and instead, we have to resort to new attacks and techniques. 

Unlike sandbox escapes and stack and heap exploitation, egghunters are an exploitation technique focused on searching memory for a specific sequence of bytes. Specifically, it uses a system call to first check whether the memory about to be accessed is valid and will not cause a segmentation fault, then checks whether that memory contains the specified bytes \cite{eggHunter}. This technique is useful as it allows a small piece of code to find and begin execution of a larger payload. However, our use of egghunters will be unique, as we will want to find interesting sequences of bytes, such as private keys and valid capabilities.

\section{Threat Model}
\label{sec:ThreatModel}

Only one threat model was considered for our attacks: an attacker with arbitrary code execution within an unprivileged library or piece of code~\cite{CheriSecAnalysis}. In particular, we aim to follow the threat model of a \textit{controlled library that does not contain implementations of the trusted computing base, such as the heap allocator or the runtime linker. It can only access particular global functions to which it has been linked.}~\cite{CheriSecAnalysis}. This means that for our threat model, we allowed our malicious library to use common functions in \lstinline{libc}, such as the allocator and the \lstinline{dlopen} function, for one of our attacks. We did not have implementations of these functions, and we could only call these functions, keeping to our threat model. In addition, system calls such as \lstinline{exec} are out of scope. 

We believe this threat model presents the most likely scenario in which a malicious actor could get code execution on the CHERI architecture, especially given the increase of high-profile supply chain attacks in recent history~\cite{ensia_supplyChain_2021}. 

For our malicious library, we assume that an attacker wants to access data either in another library or in the main binary, which they currently do not have access to due to compartmentalisation, and the restriction on what capabilities the malicious code has. This will be our main goal for our attacks: gaining access to change or read a piece of data that the trusted binary has, but which our malicious code does not. We expect our malicious library to be compartmentalised, having a distinct and isolated address space, executable code region, global offset table, and stack. This is the key mitigation and security feature that we aim to break. 

As one example, in the paper by Hammond et al~\cite{Morello_Cerise_2025}, there is a variable in memory that the malicious code is not meant to be able to read from or write to, but trusted code can. A such, this is the target of the malicious code, and will be what we aim to change when testing our vulnerabilities against this code, or code with a similar design. A similar target exists in the example code by Brodsky et al~\cite{Morello_Examples_2023}, where a key is sealed by a secure sealer capability. The key is global, but the sealing capability is only accessible to trusted code. Our aim is to get access to the sealer to then unseal the key. Our proof of concept attack has a similar goal, that being a private key generated by the OpenSSL library which is not accessible to the malicious library.


\section{Recursive Capability Scanning}
\label{sec:RecursiveScanning}

Before discussing our attacks, we show a powerful technique that proved to be very useful and successful in all of our attacks, and showcases how a malicious actor can leverage capabilities to break compartmentalisation. As the CHERI architecture requires pointers to be derived from existing capabilities, malicious code cannot create arbitrary pointers, nor can it modify pointers to arbitrary addresses. To see what memory can be reachable, we use this recursive capability scan. This technique requires a valid capability and gives the attacker a list of capabilities that it can access from this initial capability. Seeing what capabilities can be accessed from a capability is very powerful, allowing us to easily determine if we have broken compartmentalisation, and allows the malicious library to read and write to any of these capabilities. 

As a capability tells us what region of memory it can access, as well as the length of this access, we can read from the start address to the end address, and whenever we find a valid capability, we attempt to scan this new capability recursively. We additionally need to keep track of the capabilities and memory regions we have already seen, if this new capability and region of memory hasn't been seen, we add it to our data structure, and we recurse on this new capability, attempting to find new capabilities. If the new capability has been seen, we ignore it and keep looking. In essence, we aim to find the transitive closure of this initial capability, and ideally this should be an isolated closure, separate from other compartments. 

To test if a capability has been seen, we use a simple linked list. While this is not the most optimal data structure, we reason that linearly searching through memory will be significantly more intensive than linearly searching through a linked list. Furthermore, a linked list is easy to implement, which would be ideal if attempting to write this in assembly, or if trying to minimise this code, similar to what is done for egghunters~\cite{eggHunter_2004}. When testing to see if a range of memory has been seen, we test if it is a subset of an existing capability. If it is a subset, we ignore it and return. Otherwise, we move on to test this new capability against the next capability in the linked list. Additionally, we test if this new capability is a superset of an existing capability. If it is, we replace it and tell the caller to start scanning from this capability, as we have new memory regions to scan. 

This recursive process allows us to go from one capability, such as the stack pointer or a capability found in the heap, to a set of capabilities which express the transitive closure of this initial capability. The power of this technique is that it can be used on any capability with read permissions, and as such can be used on capabilities that are passed as arguments, capabilities that have been left in registers, capabilities returned by functions, or capabilities that are in any accessible memory to our malicious program.

The goal of this technique is to go from a limited set of capabilities and a compartmentalised address space, and get access to the address space of other libraries, and the main binary. If compartmentalisation is fully efficient and completely isolates libraries away from each other and the binary, this should be impossible. 

\begin{figure}
    \centering
    \begin{lstlisting}[basicstyle=\small\ttfamily,language={C}]
void scan_recursive(void* cap)
{
  for(int i = 0; i < get_length(cap); i++){
    access((char*)(cap+i), 0);
    if(errno == 14){
        // Segfault, avoiding address
        continue;
    }

    if(isValid(new_cap) && 
        getPerms(cap) == LOAD_CAP)
    {
      if(!isInList(new_cap, seenHead)){
        scan_recursive(new_cap);
      }
    }
  }
}
    \end{lstlisting}
    \caption{Code for recursive scanning.}
    \label{fig:c:scan_recursive}
\end{figure}

\section{Morello Linux}
\label{sec:Linux}

The first operating system that was analysed was Linux running on top of the Morello board, which uses the CHERI architecture. We found that it is a much weaker and less developed system than CheriBSD, as it does not have any built-in mitigations against compartmentalisation except capabilities. However, the example code provided by Brodsky et al.~\cite{Morello_Examples_2023} uses the primitives provided by the CHERI architecture to a great extent, preventing some attacks that we found that CheriBSD needed entire mitigations for. The downside of this approach is the greater emphasis on the developer writing secure code, rather than providing built-in mitigations that are turned on by default.

To determine if our exploits are effective, we use two binaries. One is a simple binary that creates an integer on the stack or heap and prints it to the console, once before calling the malicious library and once after. 
The other program is an example program included in Morello Linux. This program has an encryption key stored in a global struct, with the main defence being that this struct is sealed, preventing access unless you have the correct unsealing capability. The aim of these two programs was to have a minimal product to test the effectiveness of existing compartmentalisation models, and what the basic Linux system provides in terms of compartmentalisation. 

The example binary is compiled statically for all but one attack, and our example program is compiled dynamically.

\subsection{Vulnerabilities}

We found four vulnerabilities on this platform: stack walking, a \lstinline{dlopen} infoleak, heap scavenging, and heap storing. Each of these attacks makes use of the recursive scanning technique, and have the goal of finding capabilities that give the malicious library access outside of the compartment of the malicious library. We note that Morello Linux has less mitigations and work done on having library based compartmentalisation, which is one reason as to why these attacks are effective. 

\subsubsection{Stack walking}

The first attack we discuss is stack walking, which relies on the assumption that every program has access to the stack pointer. A malicious library can then abuse this power to read from the entire stack, including previous stack frames. Additionally, as the stack can hold capabilities, the program is able to read and use these capabilities. This is especially useful if said capabilities come from previous stack frames or compartments. Due to the potential for the stack to contain several different capabilities, we rely heavily on our recursive stack walking technique.  

\begin{figure}
    \centering
    \begin{lstlisting}[basicstyle=\small\ttfamily,language={[x86masm]Assembler}]
Capability: 0xfffff5e18314 
  [rwRW,0xfffff5e1830c-0xfffff5e18331]
Capability: 0xaaaad4fb0000 
  [rxRE,0xaaaad4fb0000-0xaaaad4fe2000]
Capability: 0xfffff5e18f20 
  [rwRW,0xfffff5e18f20-0xfffff5e19170]
Capability: 0xfffff5e1a7b0 
  [rwRW,0xfffff5e1a670-0xfffff5e1aab0]

libc base: 0xaaaad4fb0000
    \end{lstlisting}
    \caption{Results of the stack walk.}
    \label{fig:scan_results_1}
\end{figure}

Figure \ref{fig:scan_results_1} shows an abbreviated output of the recursive stack scan. To achieve this, the malicious code simply calls the recursive capability scanner shown in figure \ref{fig:c:scan_recursive} with the stack pointer. As seen in the output, if we iterate through the stack, we are able to retrieve a number of capabilities, with both read-write and read-execute permissions. Finally, the base address of the \lstinline{libc} library is printed. 

In this example, we see four capabilities being retrieved from our attack, with the first one having read and write permissions from \lstinline{0xfffff5e1830c} to \lstinline{0xfffff5e18331}. The second capability is more interesting, as it has read and execute permissions from \lstinline{0xaaaad4fb0000} to \lstinline{0xaaaad4fe2000}, and if we compare this to the address of \lstinline{libc}, either through a debugger such as GDB or in the main binary, we see that this capability has the same base address as \lstinline{libc}. As such, this allows us to call any arbitrary \lstinline{libc} function, even if it was not declared in the ELF linkage, a key security promise that CHERI aims to provide \cite{cheriDocs}. This is less useful for a malicious library as we can create these ELF linkages ourselves, but for traditional attacks that gain code execution, this would be an ideal scenario, allowing them to call any function in \lstinline{libc} and gain more control over the system. This attack additionally breaks compartmentalisation, as it allows a malicious library to access previous stack frames and data that should be isolated from it. This is a clear security issue that we see is mitigated in CheriBSD. 

In the example code by Brodsky et al~\cite{Morello_Examples_2023} (under the folder \texttt{src/compartments/privdata.c}), a key inside a global struct is sealed by a sealer, and while our malicious code has access to this global struct, it cannot use the key as it is sealed. It also cannot access the sealer as that is a privileged operation. However, by using stack scanning we are able to find the sealing capability used as it was left in an old stack frame, and by using this capability we can unseal this key. The result is shown in Figure~\ref{fig:linux_scan_results}. As seen, the trusted code seals the private data capability, and then calls the malicious code. The malicious code iterates through the stack, first looking for all capabilities with \textit{read} and \textit{write} permissions, after which it scans these capabilities for a capability with the \textit{seal} and \textit{unseal} permissions. This sealing capability is printed, and is then used to unseal the struct and get the secret, which is shown to be \textit{cafe1e55}. This attack shows that while the secret was sealed effectively, as the sealer was used in a function call, it was left in the stack from a previous stack frame and this allowed our malicious code to read this sealer. 

\begin{figure}
    \centering
    \begin{lstlisting}[basicstyle=\small\ttfamily,language={[x86masm]Assembler}]
priv:   0xfffff7efc000 
[0000fffff7efc000:0000fffff7efc060)
Sealed: 0xfffff7efc000 
[rwRW,0xfffff7efc000-0xfffff7efc060] (sealed)

Starting malicious code
starting recursive scan on stack
...
Printing list of found capabilities:
        0xfffff7eec040 
        [rwRW,0xfffff7eec040-0xfffff7eec060]
        0x22e000 
        [rwRW,0x22e000-0x242000]
        0x200000 
        [rxRE,0x200000-0x242000]
        0x2402c0 
        [rwRW,0x2402c0-0x240460]
        0x2412e0 
        [rwRW,0x2412d8-0x2416e0]
        0x20052d 
        [rR,0x20052d-0x20055a]
        0x240110 
        
looking for sealer...
finished looking through 
[rwRW,0xfffff7eec040-0xfffff7eec060]
 - found sealer: 
 0000000000000000 1 
[0000000000000000:0000000000008000) 
 G---------su------
 
sealed: 0xfffff7efc000 
[rwRW,0xfffff7efc000-0xfffff7efc060] (sealed)
unsealed: 0xfffff7efc000 
[rwRW,0xfffff7efc000-0xfffff7efc060]
secret: cafe1e55
    \end{lstlisting}
    \caption{Results of the stack walk against the example.}
    \label{fig:linux_scan_results}
\end{figure}

\subsubsection{\lstinline{Dlopen} Infoleak}

While the stack walking attack was relatively simple and relies on a capability being available to the malicious code in the stack, this attack relies on the \lstinline{dlopen} function, and as such our malicious code needs to have access to either the global offset table, a pointer to \lstinline{dlopen}, or access to the dynamic linker to work. As shown in the paper by Hammond et al.~\cite{Morello_Cerise_2025}, access to the dynamic linker and global offset table can be restricted and not given to the malicious code. We argue that as the dynamic linker is part of the trusted code base as given by Gutstein~\cite{CheriSecAnalysis}, the \lstinline{dlopen} function is in scope and to be useful in practice, an external library should be allowed to use exported functions from the dynamic linker. 

\begin{figure}
    \centering
    \begin{lstlisting}[basicstyle=\small\ttfamily,language={C}]
void dlopen_infoleak(void* cap){
    Obj_Entry* handle = dlopen("libc");
    void* mapping = handle->mapbase;
}
    \end{lstlisting}
    \caption{Pseudocode for the dlopen attack.}
    \label{fig:dlopen_pseudocode}
\end{figure}

Even though the pseudocode shown in figure~\ref{fig:dlopen_pseudocode} is simply calling \lstinline{dlopen} and saving the return value, it allows us to get a capability with read and write permissions to every loaded library, including the binary itself, as long as the code is not statically compiled. The documentation for \lstinline{dlopen} states that it \textit{returns a descriptor that can be used for later references to the object}~\cite{dlopen_man_2024}. We found that \lstinline{dlopen} actually returns an internal structure, which notably contains a capability pointing to the base of the \lstinline{.text} section of the loaded library. 

All we need to do in order to get this capability is either casting this pointer to the correct struct, or doing pointer arithmetic. The result of this attack is shown in figure~\ref{fig:dlopen_result}. As seen, the two \lstinline{pcc} registers, which control where in memory code can execute, are distinctly separate. However, we see that by calling dlopen and reading from these internal structures, we can find an executable capability to the main binary, bypassing compartmentalisation. This attack demonstrates how the stack and heap are not the only way a malicious library is able to bypass compartmentalisation, functions in the trusted code base such as dlopen may still have bugs and information leaks, and should not be assumed to be safe and secure. 

\begin{figure}
    \centering
    \begin{lstlisting}[basicstyle=\small\ttfamily,language={[x86masm]Assembler}]
In main:
my pcc: 0xaaaaaaab0bdc 
[rxRE,0xaaaaaaaa02c0-0xaaaaaaad1040]

Inside malicious library
My pcc: 0xfffff7def8a4 
[rxRE,0xfffff7ddf000-0xfffff7e10000]

l_name: ./bin/morello-dispatch
l_addr: 0xaaaaaaaa0000 
[rxRE,0xaaaaaaaa0000-0xaaaaaaad2000]
l_ld: 0xaaaaaaac0de0 
[rxRE,0xaaaaaaaa0000-0xaaaaaaad2000]
    \end{lstlisting}
    \caption{Result of the dlopen attack.}
    \label{fig:dlopen_result}
\end{figure}

Furthermore, this attack is also effective against the example compartmentalisation program, as we can retrieve the capability that points to the stack used by the binary in this internal struct. From the stack, we can then find the sealing capability and get the secret data, in the same way as the stack walking attack.


\subsubsection{Heap Scavenging}

The final attacks we describe rely on heap memory, and as such require access to an allocator. For our purposes, we use the default allocator that comes with  Morello Linux (i.e., musl libc mallocng). 

The first attack, heap scavenging, attempts to find capabilities in uncleared heap memory. By repeatedly allocating, we are able to search the heap to try and find valid capabilities. This attack has been discussed by Bramley et al.\cite{CHERI_allocator_2023}, and they describe this attack in more detail and compare it to several existing allocators in CheriBSD. We aim to show that this attack works on Morello Linux, and can be used to break compartmentalisation. 

The code in figure \ref{fig:heapScavenge} demonstrates how this can be done in  Morello Linux. As seen, this code is very simple, simply allocating several thousand times and testing if the allocated chunk has been allocated before. While the example code does get passed a pointer to a known freed address with sensitive data, we can re-purpose this code to scan each allocated chunk to try and find any valid capabilities, which can then be passed to our recursive scanner. 

\begin{figure}
    \centering
    \begin{lstlisting}[basicstyle=\small\ttfamily,language={C},showstringspaces=false]
function heap_scavenge(void* oldArray):
  for(int i = 0; i < 1000000; i++){
    int** newChunk = malloc(0x10);
    if(newChunk == oldArray){
        printf("found %#p\n", newChunk);
        printf("read value: %#p\n", *newChunk);
        printf("%d\n", **newChunk);
        **newChunk = 0x41;
    }
    free(newChunk);
  }
}
    \end{lstlisting}
    \caption{Code for the heap scavenge attack.}
    \label{fig:heapScavenge}
\end{figure}

\begin{figure}
    \centering
    \begin{lstlisting}[basicstyle=\small\ttfamily,language={C},showstringspaces=false]
void*** heapList;
int size = 220000;

int HeapStore(){
  heapList = malloc(sizeof(void*) * size);
  for(int i = 0; i < size; i++){
    void*** new = malloc(0x10);
    heapList[i] = new; 
    free(new);
  }
}

void heapCheck(void* alloc){
  for(int i = 0; i < size; i++){
    if(alloc == heapList[i]){
        printf("found %#p\n", heapList[i]);
    }
  }
}
    \end{lstlisting}
    \caption{Code for the heap store attack.}
    \label{fig:heapStore}
\end{figure}

\subsubsection{Heap Storing}


The code in figure \ref{fig:heapStore} shows the code for the heap storing attack. Similarly to the heap scavenging attack, we allocate several thousand times, but instead of checking if there are any valid capabilities in these allocated chunks, we save these pointers and free these chunks, which causes a use-after-free vulnerability to happen. Later, when the binary calls the malicious library again, or after a certain period of time, we test to see if any of the pointers in the saved list have any data written into them. If so, we know that our heap-storing attack worked, and we can now access live heap memory that the binary is currently using. 

We do note that these attacks were unsuccessful against the example compartmentalisation code, as the secret key is not stored in the heap, and neither is the sealer. Furthermore, these attacks rely on the malicious code having access to an allocator, which may not always be possible, as evidenced in the paper by Hammond et al~\cite{Morello_Cerise_2025}. Additionally, this attack relies on the program allocating heap memory and storing capabilities inside it without clearing or overwriting it before freeing. This may not always be the case, and even if capabilities were stored on the heap, they may not lead to a full compartment escape as they may only point to small subsets of memory. Because of these limitations, we believe heap scavenging and heap storing are not as fruitful or useful as stack walking and the \lstinline{dlopen} infoleak.

\subsection{Mitigations}

As these four attacks are relatively simple, we propose several mitigations that would be effective at stopping these attacks. 

For stack walking, there are several approaches you can take. Clearing sensitive capabilities from the stack would work well, but you would have to be careful that your zeroing is not optimized away by compiler, and that you always remember to do so. Creating an independent stack for each compartment is another effective mitigation, and is what CheriBSD has done with their c18n mitigation~\cite{CHERI_c18n_2024}. 

For the \lstinline{dlopen} infoleak, we propose sealing the handle in \lstinline{dlopen} before returning it. We believe that this option is the best one at preventing this class of vulnerability as it uses existing primitives of the architecture and should not have any impact on existing code, as this is an internal struct and should not be interacted with. The only potential issue with this solution is the issue of securely storing the keys used for sealing, and making sure that this capability is unsealed when it is used in functions such as dlsym. If the keys are improperly stored, a malicious library could easily unseal the returned struct, bypassing this mitigation, as was shown in the stack walking attack against the Morello Linux compartmentalisation example. 

For heap scavenging and heap storing, we propose that the current allocator could use a separate memory space for each compartment, but this will significantly increase the memory overhead, especially if a program has several compartments. The issue with this approach is that there is the possibility of leaking a capability to these isolated heap arenas to other compartments. The second mitigation proposed is to zero memory before returning it to the program to prevent the heap scavenging attack, and to use a mitigation called heap revocation from CheriBSD~\cite{CHERI_temporal_safety_2024} to prevent the heap storing attack.

However, if you only need to protect one piece of data from a malicious library, you may not even need compartmentalisation. The example program provided in  Morello Linux shows how sealing can be used to protect a piece of data. By using specific instructions to unseal the data before passing it to a function, it remains sealed by default, protecting it even if a malicious library were able to bypass compartmentalisation. However, you would still need to protect the key from the malicious library, as otherwise the malicious library can simply find the key and unseal the protected data by itself. 

We see that these attacks all have a similar theme: current isolation enforcement techniques are not enough to fully compartmentalise malicious code, and more work needs to be done to achieve compartmentalisation.

\section{CheriBSD}
\label{sec:CheriBSD}

CheriBSD is the second operating system analysed, with it being a port of BSD onto the CHERI architecture. While similar to Morello Linux, it has several mitigations that are not present in Morello Linux, and is receiving updates and security fixes more often. The main mitigations that we will focus on are c18n \cite{cheri_c18n_2023}, a security feature focused on confining compartments to their own stack space, and heap revocation, which aims to prevent use-after-free attacks \cite{CHERI_heap_revokation_2023}. These two mitigations, especially c18n, provide much more isolated compartments that malicious libraries would have to break out of. 

\subsection{Vulnerabilities}

The same four vulnerabilities exist to some extent on CheriBSD. Naturally, the different operating systems and the two security features \lstinline{c18n}, and heap revocation, do have an effect on the vulnerabilities. We will discuss what, if anything, needs to be changed to make these attacks work on CheriBSD, what these attacks give an attacker, and how effective they are at breaking compartmentalisation. 

\subsubsection{Stack walking}

Running the same code with some edits on CheriBSD gives us the result seen in figure \ref{fig:scan_results_2}. While we do get a similar result and are able to find several new capabilities, notably the executable capabilities are sealed, preventing us from using them to get arbitrary code execution, as these capabilities are immutable. We do still get a capability that points to the start of the \lstinline{libc} library; however, unlike Morello Linux, it has read and write permissions, instead of read and execute. This means we cannot call arbitrary functions and may not be able to make as much use of this as we could in Linux, but we can potentially still edit internal \lstinline{libc} variables, such as function hooks and internal \lstinline{libc} data.

\begin{figure}
    \centering
    \begin{lstlisting}[basicstyle=\small\ttfamily,language={[x86masm]Assembler}]
Capability: 0xfffffff7b850 
    [rwRW,0xffffbff80000-0xfffffff80000]
Capability: 0x40152d61 
    [rxR,0x40130000-0x40190700] (sentry)
...
Capability: 0x40389521 
    [rxR,0x40209000-0x40871000] (sentry)
Capability: 0x40195b40 
    [rwRW,0x40195b40-0x40195ba0]
...
Capability: 0x40222f38 
    [rwRW,0x40209000-0x40871000]
Capability: 0xfffffff7c9f0 
    [rwRW,0xfffffff7c9f0-0xfffffff7ca20]
...

(gdb) info proc map
Start Addr    End Addr    Flags      File
0x40209000   0x402bb000  r-- CN--  /libc.so
    \end{lstlisting}
    \caption{Result of the stack scan in CheriBSD.}
    \label{fig:scan_results_2}
\end{figure}

However, the \lstinline{c18n} mitigation fully prevents this attack from working, creating fully isolated stacks for each library. Even with the recursive stack walking technique, we were unable to find any capabilities outside of this malicious compartment. The result of this mitigated attack is shown in Figure~\ref{fig:scan_results_c18n}, and as seen all of the capabilities are either sealed, or point to this isolated stack. This mitigation highlights the effectiveness of the sealing primitive, and how if can be used effectively for compartmentalisation. While the \lstinline{c18n} mitigation is effective at preventing stack walking, it is not a perfect compartmentalisation primitive, but we acknowledge that it is still a work in progress.

\begin{figure}
    \centering
    \begin{lstlisting}[basicstyle=\small\ttfamily,language={[x86masm]Assembler}]
In main, Stack capability: 
0x41fffef0 [rwRW,0x41c00000-0x42000000]

In library, Stack capability: 
0x427fff70 [rwRW,0x42400000-0x42800000]

Found capabilities:
    0x41419000
[rwRW,0x41419000-0x41419020]
    0x409c3d21
[rxR,0x409c3c50-0x409c3d98] (sentry)
    0x427ffec0 
[rwRW,0x42400000-0x42800000]
    0x409c3dcd 
[rxR,0x409c3da0-0x409c3ee8] (sentry)
    0x409c41bd
...
    0x40195cc0 
[rwRW,0x40195cc0-0x40195d20] (sealed)
    0x4038a751 
[rxR,0x4020a000-0x40873000] (sentry)
    0x401cd1a0 
[rxR,0x4019c000-0x401ce000]
    \end{lstlisting}
    \caption{Result of the stack scan in CheriBSD with c18n enabled.}
    \label{fig:scan_results_c18n}
\end{figure}

\subsubsection{\lstinline{Dlopen} Infoleak}

Similarly to Morello Linux, \lstinline{dlopen} returns a \lstinline{void*}, which in reality is a pointer to a \lstinline{Obj_Entry} struct. This struct is different to the struct returned in Morello Linux, but by doing some pointer arithmetic, we can get the \lstinline{mapbase} to the library in a similar manner, allowing us to read and write to the \lstinline{.text} section. Not only can we get the base of any library, we only need to call \lstinline{dlopen} once, as the struct is a doubly linked list, allowing us to get the mapping of every library that is dynamically loaded, including the binary itself. We can then use our recursive scanning technique to find all of the capabilities that the binary is using, further expanding the reachable address space of our malicious library.

\begin{figure}
    \centering
    \begin{lstlisting}[basicstyle=\small\ttfamily,language={C}]
typedef struct Struct_Obj_Entry {
  /* Magic number (sanity check) */
  Elf_Size magic;		
  /* Version number of struct format */
  Elf_Size version;		

  TAILQ_ENTRY(Struct_Obj_Entry) next;
  /* Pathname of underlying file (\%) */
  char *path;			
  /* Directory path of origin file */
  char *origin_path;			
  /* Count of transient references */
  int holdcount;		
  /* Number of times loaded by dlopen */
  int dl_refcount;		
  /* Base address of mapped region */
  /* This includes the .text section and more */
  caddr_t mapbase;		
  /* Size of mapped region in bytes */
  size_t mapsize;		
  /* Base address in shared object file */
  Elf_Addr vaddrbase;		
  // more entries follow
}
    \end{lstlisting}
    \caption{Internal Obj\_Entry Struct in CheriBSD.}
    \label{fig:objEntry}
\end{figure}

If we look at the \lstinline{Obj_Entry} struct in Figure \ref{fig:objEntry}, we see that it stores a capability to the entire \lstinline{.text} section in the \lstinline{mapbase} field, and internally, \lstinline{dlopen} returns a pointer to an \lstinline{Obj_Entry} struct, but cast as a void pointer, which is why we can access all of these values. The code in Figure \ref{fig:c:dlopenAttack} demonstrates how this \lstinline{Obj_Entry} struct can be manipulated to get the capability to every loaded library. From this, we can then access the \lstinline{mapbase}, a capability that points to the base of the \lstinline{.text} section and covers the entire library. 

\begin{figure}[h!]
    \centering
    \begin{lstlisting}[basicstyle=\small\ttfamily,language={C},showstringspaces=false]
function exploit_dlopen():
  void* lib = dlopen("libc.so", RTLD_NOW);
  Obj_Entry* obj = (Obj_Entry*)lib;

  Obj_Entry* obj_next = TAILQ_NEXT(obj, next);
  do{	
    printf("Map base: %#p\n", obj_next->mapbase);	
    obj_next = TAILQ_NEXT(obj_next, next);
  } while(obj_next != NULL);
}
    \end{lstlisting}
    \caption{Code for the dlopen attack}
    \label{fig:c:dlopenAttack}
\end{figure}

\subsubsection{Heap Scavenging}

Heap scavenging is just as effective in CheriBSD as it is in Morello Linux, and allows for a malicious library to gain capabilities to memory outside of its compartment. However, we note that this attack is the same as the attack found by Bramley et al~\cite{CHERI_allocator_2023}, and is found to be ineffective against the \lstinline{snmalloc} allocator. However, we aim to show that this attack can be used to bypass compartmentalisation between a library and a binary. 


\subsubsection{Heap Storing}

Unlike in Morello Linux, heap storing is ineffective due to the malloc revocation shim. This mitigation places each freed chunk of heap memory into a quarantine, and after a certain epoch it invalidates any capabilities that point to this freed heap memory. Because our attack stores pointers to freed heap memory, after this epoch and before this heap memory is given back to the binary, our capabilities are invalidated.

\subsection{Mitigations}

The mitigations proposed for CheriBSD are similar to those proposed for Morello Linux. The existing mitigations in CheriBSD, \lstinline{c18n} and heap revocation already mitigate stack walking and heap storing effectively and should be used. 

For stack walking, we know that the c18n mitigation is effective, as it prevented our stack walking attack and is very effective at doing so. It is still experimental, and needs to be tested for security and performance much more than in this paper \cite{CHERI_c18n_2024}. The other three recommendations have not been tested, and remain theoretical as we found c18n to be by far the best mitigation against this attack. 

To mitigate the \lstinline{dlopen} infoleak, we propose the same protection that was described fro Morello Linux; sealing the handle with a key only known to the dynamic loader. This would prevent a malicious program from abusing the struct to escape its compartment, but would require more bookkeeping, as the struct would have to be unselaed whenever it is passed to a function such as dlsym. We believe that sealing would be effective, as it would make the handle immutable and truly transparent to the developer, preventing them from doing low level pointer arithmetic to find capabilities.

To mitigate heap scavenging, we propose two mitigations. The first being that a call to free should zero memory before it can be reallocated. This would impact performance, as it would make each call to malloc equivalent to a call to calloc. The other mitigation would be to use two allocators, one safe allocator used by trusted compartments, and one unsafe allocator used by untrusted code. This mitigation would impact performance and memory usage, but as we have shown that any capabilities stored on the heap can be retrieved by any compartment, we believe that this mitigation would be useful. 

All in all, the mitigations present in CheriBSD are verry effective, and mitigate two of our attacks, which is a very good sign for the maturity of this platform.

\section{Proof of Concept}
\label{sec:POC}

Lastly, we show what breaking compartmentalisation can achieve in a proof-of-concept program, where we gain control of a private key generated by the OpenSSL library and stored in the binary, which is not given to the malicious library. The code for this example program is shown in figure \ref{fig:c:sslExample} \cite{stackoverflow_ssl}.

\begin{figure}
    \centering
    \begin{lstlisting}[basicstyle=\small\ttfamily,language={C}, showstringspaces=false]
int main(){
  printf("in main, creating private key..");

  const int kBits = 1024;
  const int kExp = 3; 

  RSA *rsa = RSA_generate_key(kBits, kExp);

  printf("rsa: %s (%#p)\n", rsa, rsa);

  // call malicious library
  malicious_library();	
}
    \end{lstlisting}
    \caption{Code for the binary in the proof of concept attack}
    \label{fig:c:sslExample}
\end{figure}

We see that the code creates a private key from the OpenSSL library and then calls our malicious library. Ideally, compartmentalisation should prevent the malicious library from accessing this private key and any other data in the binary \cite{hardware_cap_compartmentalization_2023}, but by using the attacks discussed above, we show how we are able to read this private key. The code for the malicious library is shown in Figure~\ref{fig:c:sslExploitExample} and is dynamically compiled together in an isolated compartment with the binary.

\begin{figure}
    \centering
    \begin{lstlisting}[basicstyle=\small\ttfamily,language={C},showstringspaces=false]
void malicious_library():
  void* object = dlopen("libc.so.7", RTLD_NOW);
  Obj_Entry* obj = (Obj_Entry*)object;

  obj += 6;
  void* map = *obj;

  printf("testing strings...\n");
  char* testString = "BEGIN RSA PRIVATE KEY";
  for(int i = 0; i < 0x200000; i++){
    char* testKey = map+i;
    if(strlen(testKey) < 100)
        continue;
    if(strncmp(testKey, 
        testString, 
        strlen(testString)) == 0)
        printf("%s\n", testKey);
  }
  return;
}
    \end{lstlisting}
    \caption{Code for the malicious library in the proof of concept attack.}
    \label{fig:c:sslExploitExample}
\end{figure}

\begin{figure}
    \centering
    \begin{lstlisting}[basicstyle=\small\ttfamily,language={[x86masm]Assembler},showstringspaces=false]
(gdb) p pem_key
$1 = 0x41579000 [rwRW,0x41579000-0x41579380] 
"-----BEGIN RSA PRIVATE KEY-----"

gef>  xinfo pem_key
xinfo: 0x41584000 
Page: 0x0000000041400000  ->  0x0000000041600000 (size=0x200000)

Permissions: rw-
Pathname: [heap]
    \end{lstlisting}
    \caption{Locating the private key in GDB.}
    \label{fig:c:gdbDumpSSL}
\end{figure}

The malicious program first uses \lstinline{dlopen} to create a handle to any library, in this case the \lstinline{libc} library, as we assume it will always be available, and as such \lstinline{dlopen} will always return successfully. Then, we use the extra data returned with the handle to iterate over the linked list with every loaded library in this program, as was shown in the \lstinline{dlopen} infoleak attack. We know that the private key is located at a certain region of memory by using a debugger such as GDB and printing the location of this key. Figure~\ref{fig:c:gdbDumpSSL} shows that this private key is located in one of the sections in the binary, which is outside of the compartment of our malicious library. We can then use our recursive scanning technique to attempt to find a capability that has this address within its bounds. Even though we know roughly where the private key is located, due to address space randomisation, this offset is random every time the program is run, so we still need to find it within the correct section. Luckily, this is easy, as this private key contains a header which we can scan for. Finally, we print any private keys located to verify that they are the same as the ones generated by the binary. The full output for our attack is shown in \ref{fig:c:sslScanAttack}. As seen, the \lstinline{libssl} library did not have the capability to section, and as such we had to look through other libraries. Instead, a capability to the section we want was found in \lstinline{libthr}, as well as 74 other capabilities. This indicates a potential bug or issue in \lstinline{libthr}, as it is a library that provides POSIX threads and contains far more capabilities than any other library.

\begin{figure}
    \centering
    \begin{lstlisting}[basicstyle=\small\ttfamily,language={[x86masm]Assembler},showstringspaces=false]
path: /usr/lib/libssl.so.30
Found: 
  0: 0x4151b480 
  [rwRW,0x4151b480-0x4151b4a0]
  1: 0x4014ac11 
  [rxR,0x40130000-0x40190700] (sentry)
  2: 0x40195b40 
  [rwRW,0x40195b40-0x40195ba0]
...
path: /lib/libthr.so.3
Found: 
  0: 0x41554140 
  [rwRW,0x41400000-0x41600000]
  1: 0x40195840 
  [rwRW,0x40195840-0x40195880]
  2: 0x4103e000 
  [rwRW,0x41032000-0x4103e000]
  ...
  70: 0x40152d61 
  [rxR,0x40130000-0x40190700] (sentry)
  71: 0xffffbff80000 
  [rwRW,0xffffbff80000-0xfffffff80000]
  72: 0x4141e000 
  [rwRW,0x4141e000-0x4141ea00]
  73: 0x40357521 
  [rxR,0x401d7000-0x4083f000] (sentry)
  74: 0x40f376c5 
  [rxR,0x40f17000-0x40f82000] (sentry)

Found: 0x41400000 [rwRW,0x41400000-0x41600000]
testing strings...
0x41582000 [rwRW,0x41400000-0x41600000]:
-----BEGIN RSA PRIVATE KEY-----
    \end{lstlisting}
    \caption{Capabilities and private key found through recursive scanning.}
    \label{fig:c:sslScanAttack}
\end{figure}

This attack is significantly easier as the private key is encoded as a string and has a detectable header. If the private key were kept as a raw binary representation, our attack would be harder to achieve, but it should still be possible, as the malicious library would know roughly how large the memory footprint of the private key is and could search for capabilities with that length. Another limitation of this attack is that it requires knowing the general location of the private key, as you cannot always assume that the memory layout of every program will be the same. 

We see that even though compartmentalisation is active, and initially isolates the malicious library, we can use \lstinline{dlopen} to get a capability to an internal struct, which has capabilities that point outside this isolated compartment. This is the key weakness with library based compartmentalisation when compared to traditional IPC compartmentalisation. Even a single leak of a capability can lead to a malicious library bypassing compartmentalisation and being able to access almost the entire address space of the binary.

\subsection{Mitigations}

As seen in the Morello Linux examples, sealing data is one of the best ways of protecting data, even from attacks that have no current mitigations. Sealing does have the issue of secure key storage, but as long as you clear the key from easily accessible memory, such as the stack, it should be secure. This security feature could be implemented in this proof of concept by simply sealing the generated private key.

We believe that if implemented correctly, sealing the private key would be an effective mitigation against this proof-of-concept attack. The only issue with this mitigation would be the secure storage of the key, as storing it in widely accessible memory would simply add an extra step for the attacker, that step being finding this key. 

However, we note that the dlopen attack was able to find a second copy of the private key, so mitigating the \lstinline{dlopen} infoleak would also stop this proof of concept attack from being effective, as that is the key attack that we use in order to get the private key.

\section{Discussion and Conclusion}
\label{sec:Discussion}

We aim to show four different attacks that allow a malicious library to escape an isolated compartment, and gain access to memory that it normally should not be able to access.
We see that all four attacks worked on Morello Linux, and two of them also worked on CheriBSD with both the \lstinline{c18n} and heap revocation mitigations present. These attacks were also relatively simple, being very easy to achieve from a malicious library and granting the library a means of escaping its compartment. Due to this, we propose that the mitigations outlined be implemented. Furthermore, some of the claims of compartmentalisation fall short due to bugs and information leaks in the dynamic loader, the stack, and memory in the heap. However this is not an issue with the architecture; rather, it is an issue in the operating systems, the compilers, and the dynamic linkers that are being used in CheriBSD and Morello Linux. We also recognise that compartmentalisation is still being developed and worked on, and we hope that these attacks illustrate some area for improvement.  

These vulnerabilities rely heavily on information leaks, which become much more significant in the CHERI architecture, as any pointer reveals its validity, address range, permissions, and other metadata that is very useful for an attacker. We discovered a basic attack primitive that makes use of this additional metadata, that being the recursive scan from a starting capability, which allows a malicious capability to see what parts of memory it can reach. This basic primitive had success against the stack, but due to the \lstinline{c18n} security feature of CheriBSD, this is not fully effective on that operating system. However, Morello Linux was vulnerable to the recursive stack scan as it has no stack isolation. The next vulnerability we found was that the \lstinline{dlopen} function returned an internal struct to the user, which contained several capabilities that gave the malicious code extra capabilities, which were much more permissive than they should have been. These extra capabilities allowed a malicious library to read to several different sections of each loaded library, as well as the main binary itself. Finally, it was discovered that the memory allocators in both systems were vulnerable to scavenging attacks, those being attacks that looked for valid capabilities in heap memory that was freed by another compartment. This memory is not cleaned unless explicitly asked to, which is the main vulnerability exploited. Heap storing was also explored, which relies on the malicious library keeping a pointer to a freed chunk of memory and reading from this memory at a later date, but CheriBSD has an effective mitigation against this, that being the \lstinline{malloc revocation shim}. In total, two of the vulnerabilities, the \lstinline{dlopen} infoleak and heap scavenging, are fully effective against both systems given access to certain shared libraries and functions. The other two attacks, stack walking and heap storing, are effective but are more limited due to mitigations. 

It was found that while there was literature about how CHERI provides security mitigations against common memory corruption bugs, there was not enough research on the compartmentalisation that CHERI can provide, especially for compartmentalising potentially malicious libraries. In this paper, we attempt to understand the effectiveness of compartmentalisation when faced with a malicious library, and how such a library would attempt to break compartmentalisation. From this research, we discovered that both CheriBSD and Morello Linux have vulnerabilities that allow a malicious library to break compartmentalisation. These vulnerabilities were stack walking, the \lstinline{dlopen} infoleak, heap scavenging and heap storing. The attacks allowed a malicious library to escape its compartment and gain access to memory that it normally should not have access to. 

In conclusion, while the CHERI architecture provides developers with capabilities that protect against classic memory corruption attacks, there are still ways in which these capabilities can be used in buggy or insecure ways, allowing a malicious library to break out of an isolated compartment. 

\bibliographystyle{IEEEtran}
\bibliography{references}

\end{document}